\documentclass[
]{ceurart}

\usepackage{listings}
\usepackage{booktabs}
\usepackage{multirow}
\usepackage{tabularx}

\usepackage{etoolbox}

\newtoggle{longversion}
\settoggle{longversion}{true}
\newcommand{\longonly}[1]{\iftoggle{longversion}{#1}{}}

\usepackage{hyperref}

\begin{document}

\copyrightyear{2026}
\copyrightclause{Copyright @ 2026 for this paper by its authors. Use permitted under Creative Commons License Attribution 4.0 International (CC BY 4.0)}

\conference{Joint Proceedings of the STAF 2026 Workshop, LLM4SE Rennes, France, June 29- July 3, 2026}

\title{How to Compare the Security of Code Written by Humans to LLM-generated Code}


\author[1]{Jasmine Egli}[orcid=0009-0002-0160-5709]
\author[1]{Rebecca Balebako}[orcid=0000-0001-9862-0790]
\cormark[1] 

\address[1]{Zurich University of Applied Sciences}

\cortext[1]{Corresponding author.}

\begin{abstract}
Large language models (LLMs) are rapidly transforming how software is created and maintained. Comparing LLM-generated code against human-written standards is essential to determine whether these new tools uphold or erode the security baselines established by professional developers. Yet, we lack a standardized method for empirically comparing the security of code produced through human-LLM collaboration against LLM-only, or traditional human-only methods. To facilitate this, we propose an automated framework for conducting comparative studies across human-only, LLM-only, and hybrid conditions. Our approach automates the logging of prompts, timing, and experimental settings, measuring outcomes through multi-dimensional static and dynamic quality analysis. We provide an open-source implementation of this framework to ensure that future researchers can conduct reproducible, species-fair experiments. Importantly, we validate the framework via a feasibility study, providing an experimental blueprint for ``species-fair'' comparisons between human and AI subjects.  By sharing lessons learned, we establish a foundation for empirical research on human and LLM-generated code for software security.
\end{abstract}

\maketitle

\section{Introduction}
\label{Introduction}

Large language models (LLMs) have fundamentally altered software development. While LLM tools offer significant efficiency gains , security remains a primary, unresolved concern \cite{ramirez_state_2024}. Current literature offers disparate conclusions on the safety of LLM-generated code, often because outcomes are artifacts of inconsistent evaluation methods and contexts.
 
There is relatively sparse literature on how humans and LLMs compare in producing secure code. To limit confounding factors, comparing human and LLM-written code requires a ``species-fair" approach \cite{firestone_performance_2020}.  Empirical research that attempts to compare the security of LLM versus human code should follow best practices in Developer Centered Security research and best practices in LLM evaluation, while striking the balance in what is a fair comparison.  

Furthermore, evaluations of human and LLM-written code are often confounded by non-deterministic outputs, varying task constraints, and a reliance on subjective or high-friction manual reviews. Without a reproducible and multi-dimensional measurement pipeline, it is difficult to determine whether observed security flaws are merely artifacts of inconsistent experimental design.  To address these inconsistencies, future empirical work must employ precise, species-fair designs that account for these influences.  There is an urgent need for an automated, deterministic framework that can bridge this gap and enable researchers to isolate the variables that influence code security across human, LLM, and hybrid generation paradigms

Our developed framework proposes a path to bridging the gap.  It offers researchers a method to compare the functionality and security of human-written code and LLM-generated code. This framework can help designers, security researchers, and tool developers identify conditions and variables that influence code security across different code generation paradigms.  By providing humans and LLMs with identical exercises and executing the results in identical environments, and then utilizing code quality metrics to understand code security, the framework allows for a direct assessment of how human logic compares to the capabilities of modern models.

We evaluate the framework on curated human code and LLM-generated code to demonstrate the feasibility of this approach. We provide lessons learned for experimental design, task creation, and security metrics to ensure reproducibility. Our contribution lies in the identification of experiment design issues, providing a roadmap for future researchers to align LLM secure programming capabilities with commensurate human skill levels in a way that is scientifically defensible.
\begin{enumerate}
    \item Instructional Symmetry: Identifying the prompt engineering necessary to maintain a fair comparison between human and LLM instructions.
    \item Model Reliability: Documenting how specific model behaviors can break automated evaluation pipelines.
    \item Analysis of Experimental Attrition: Categorizing the "leaky pipeline" where samples are lost to malformed outputs or execution failures, providing a blueprint for the sample over-provisioning required for statistical validity.
\end{enumerate}

The next sections are as follows.  In the \nameref{RelatedWork} section, we give an overview of relevant work in evaluating secure code.  In the \nameref{MethodFramework} section, we describe the design decisions made in designing the framework tool and outcomes.   In the \nameref{FSMethods} we describe the feasibility design and lessons learned.  Finally, in the section \nameref{Discussion}, we cover limitations, future work, and ethics.  

\section{Related Work}
\label{RelatedWork}

\subsection{Measuring the security of code}
Despite its complexity, quantifying security is essential for comparative code evaluation. The focus of this work is on source code, as opposed to broader systems of security, as this is an area LLMs are currently helpful. 

\longonly{Secure code evaluation cane be described in three rough groups: 1. Researchers manually review and rate code \cite{tahaei_survey_2019} \cite{acar_you_2016}. This is not only labor-intensive but requires checks for reliability.  2.  LLMs act as judges to evaluate the security of code, which may introduce hallucinations or other errors.  3.  Use tools such as static analysis to evaluate code for specific metrics, which can introduce the bias of any specific tool into the evaluation \cite{dai_rethinking_2025}.  To focus on deterministic, repeatable outputs, we use focus on the third option.}

\subsubsection{Run-time Analysis for Errors and Correctness}
\citet{dai_rethinking_2025} examine LLM-generated code, and show the need to measure both functionality and security of code.  Correctness include error rates and functionality. Code should run without failure and it should complete the required task.  

Frequent runtime errors can indicate fragile code. High error rates during execution suggest a lack of robust input validation, which is the primary entry point for injection attacks. \citep{shin_empirical_2008} demonstrated that traditional software fault metrics are strongly correlated with security vulnerabilities, suggesting that ``buggy'' code is statistically more likely to be ``vulnerable'' code.  

\longonly{Fuzzers are an example of run-time analysis \cite{manes_art_2021}.  By running the program with various inputs, they measure whether the programs fail or produce warnings.  This can help find bugs and vulnerabilities in specific contexts.  Fuzzing is a reasonable extension of this work.}

\subsubsection{Static Analysis for Code Quality}
Static analysis automates code evaluation against a predefined rule set, enabling scalable, fast, repeatable quality checks. However, static analysis cannot include contextual nuances, runtime behavior, and emergent properties. Despite these limits, static analysis integrates into automated pipelines and has relevance to code security \cite{pistoia_survey_2007,choi_static_2021,gosain_static_2015}.  
Key metrics include \textit{Cyclomatic Complexity:} (McCabe), which measures the number of linearly independent paths through a program's source code and acts as predictor of security vulnerabilities \cite{shin_empirical_2008}.  \textit{Mean function length:} or lines of code per function is another indirect measure of security. Long functions tend to ``leak'' state, where variables intended for one task are accidentally reused for another, leading to memory corruption or sensitive data exposure \cite{shin2008empirical,moshtari2013using,tehrani2024assessing}.

\longonly{
\subsubsection{Manual Analysis}
Manual analysis is the process where human experts meticulously inspect source code to identify security flaws, logic errors, and violations of secure coding standards. Unlike automated tools that rely on predefined patterns, manual analysis leverages human understanding of the application's business logic to uncover "context-aware" vulnerabilities \cite{charoenwet_toward_2024}. Manual analysis has pros and cons, with some research finding the effectiveness to be highly variable \cite{edmundson_empirical_2013}.  While not a part of our framework, it would be possible to additional include manual evaluation scores in the final assessment. 

\subsubsection{Developer Experience}
Some studies on code security focus on human experience, such as time to completion \cite{acar_comparing_2017} or self-reported confidence in the code\cite{perry_users_2023}.  Longitudinal work has also included total time to develop and review the LLM-assisted code generation \cite{becker_measuring_2025}.  Our framework does not include developer experience metrics, as we focus on LLM-generation and automated measurement.   
}

\subsection{Developer Centered Security}
Code security is shaped by environment in which the code is produced. Developer-Centered Security (DCS) is a subset of the usable security community.  This field include over a decade of work to understand what conditions influence secure code, and offer insights into the contexts in which developers produce secure (or insecure) code \cite{tahaei_survey_2019} \cite{acar_you_2016}.

\longonly{
DCS studies generally use one of three experimental methods.  Qualitative research can include interviews or small surveys (including \cite{balebako_privacy_2014, braz_software_2022, gutfleisch_how_2022, assal_think_2019}).  Observational research involves examining existing repositories \cite{meli_how_2019, fischer_stack_2017}. Experimental studies ask developers to write code with specific functional requirements under specific conditions.  In some cases, developers are asked to come to the lab, and in others they can perform the task remotely (for example \cite{stransky_lessons_2017}). Recruiting software developers is particularly challenging for software engineering research\cite{alami_are_2024, wang_end-users_2024}, and usable security is no exception\cite{kaur_where_2022, tahaei_lessons_2022}. Because developer recruitment is a significant bottleneck, our work provides the necessary foundation to identify and resolve experimental flaws before engaging human subjects.

Recruiting software developers is particularly challenging for software engineering research\cite{alami_are_2024, wang_end-users_2024}, and usable security is no exception\cite{kaur_where_2022, tahaei_lessons_2022}.  Participants for studies can be drawn from freelance platforms, crowdsourcing, recruitment in developer communities, or directly within cooperative companies \cite{serafini_recruitment_2023}. As recruitment is already challenging, our framework strives to ensure that code output from each developer measured in a high-quality, reproducible manner. 
}

Current empirical evaluations of LLM-assisted coding in DCS have yielded fragmented results, with findings on security and productivity varying significantly across different study designs.  \citet{sandoval_lost_2023} used a randomized control trial to understand how novice programmers were impacted by LLMs.  Overall, they found LLMs assistants had small ``but consistent'' benefits to functionality of the code without adversely impacting the security of the code.  \citet{perry_users_2023} discovered that participants with access to an LLM assistant wrote less secure code than those without access to an assistant. \citet{belozerov_secure_2025} explored the capabilities of ChatGPT in detecting security issues but concluded that LLM-generated code tends to have more vulnerabilities than human-written code. \longonly{\citet{becker_measuring_2025} conducted a randomized controlled trial to understand how LLM tools affect the productivity of experienced open-source developers and found that, contrary to expectations, the completion time of development tasks actually increased by 19\%.}

\subsection{Comparing Human to LLM code}
Direct comparisons of code by humans and by LLMs are often confounded by asymmetrical testing conditions, necessitating a more balanced approach to evaluating human and LLM-generated software. \citet{firestone_performance_2020} advocates for ``species-fair" evaluations, emphasizing the necessity of equivalent constraints when benchmarking human and machine performance. Drawing parallels from computer vision, the author argues that direct assessments are only valid when both agents operate under symmetrical environmental and task-based parameters.  

\longonly{Prompt engineering is the practice of refining the input queries of "prompts" to LLMs to meet certain goals \cite{marvin_prompt_2024}. \citet{bruni_benchmarking_2025} suggests the prompts can have significant impact in the security of LLM-generated code, as measured by static analysis for specific vulnerabilities.  State-of-the-art methods for secure-code generation from LLMs describe fine-tuning models with secure and insecure code\cite{he_large_2023, he_instruction_2024}.  A comparable training method for humans may include a comprehensive training or a course on secure code.  Other secure code generation methods include reasoning loops in which the model is prompted to re-think their code with security in mind, perhaps based on results from a static analyser \cite{nijkamp_codegen_2023}.  Similar tasks for humans would likely include multiple steps in which coders are specifically asked to review their code with security in mind.   }

Previous research on comparing humans and LLMs have not relied on species-fair comparisons.  \citet{molison_is_2025} integrated Python solutions from open source repositories to test various LLM configurations to measure maintainability, reliability, and remediation effort. However, the LLM prompts included zero-shot, few-shot, and fine-tuning.  Arguably this would be species-fair if humans also had multiple attempts to write the code.  \citet{cotroneo_human-written_2025} analyzed over 500,000 code samples in Python and Java to compare human developers to LLMs, also using static analysis for code security and measuring code complexity. The LLMs were prompted to generate code by giving them the docstring and function signature that was extracted from the human code.  This arguably means the LLMs started with a different task than the humans.  \citet{licorish_comparing_2025} used the same prompt for humans and LLMs, therefore being more ``species-fair.'' With a dataset of 72 software engineering tasks, they used static analysis tools  to measure coding standards, complexity, and security vulnerabilities.  However, only one human was responsible for the full human-generated code dataset, which introduces concerns about representativeness of human coders. The findings across these papers are contradictory in regards to whether LLMs or humans are "better" or "more secure", and some were task dependent. 

\longonly{A further issue emerges in LLM to human code comparison.  Researchers analyzing human and LLM generated code must contend with noise in the experimental data.  For example, both LLMs and humans are inherently non-deterministic, meaning identical tasks and setups can yield various code results, even from the same LLM or human\cite{astekin_exploratory_2024, cui_language_2025, sawadogo_revisiting_2025}.  Work systemizing such developer centered security study research has provided information to estimate the statistical power of developer surveys \cite{ortloff_sok_2023}.  Research comparing humans to LLMs will need to determine also the power of any LLM-generated code samples, due to  residual nondeterminism in LLM internals \cite{beckers2025large} \cite{song2025good} \cite{ouyang2025empirical}.  }

\section{Framework Implementation}
\label{MethodFramework}

\subsection{Creating and Collecting Code}
Comparing human-created code to LLM-generated code requires acquiring code from humans and prompting LLMs to generate code. Researchers may ask participants and LLMs to write code for different tasks. We refer to these  human-language instructions as "exercises."  For example, an simple exercise might be to ``Write python to generate the first 10 values in Fibonacci sequence of numbers.''  For LLMs, the exercise is often called a ``prompt.''  Our framework is designed to track and compare results for multiple exercise definitions, and assume the same exercise is given to both humans and LLMs.  After a human or LLM receives the exercise, they should generate executable code.

\subsection{Framework Design}
\begin{figure*}[h]
\begin{center}
    \includegraphics[width=\textwidth]{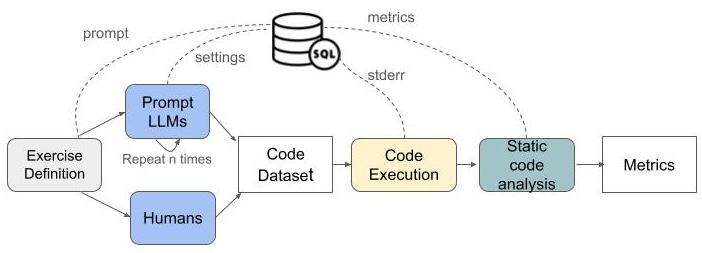 } 
    \caption{Framework design: Exercise definition as prompt for LLMs and humans, the output is executed in a container, and static analysis is run.  All settings and environment variables are saved the database.}
    \label{fig:evaluation_pipeline}
\end{center}
\end{figure*}

The architecture overview of the implementation is displayed in Figure\ref{fig:evaluation_pipeline}.  In general, this framework resembles many of the related work examples, with additional emphasis in reproducibility and species-fair design.  The framework serves a dual purpose: first, as a standardized interface for code generation across various LLMs, and second, as a deterministic execution and evaluation pipeline for security measurement. Key properties of a deterministic system include: (1) clear definitions of input and output variables, (2) precisely known initial conditions, (3) cause-effect relationships, (4) predictability, (5) continuity, (6) reproducibility, (7) modeling, and (8) time independence. This architecture ensures that our measurements of code quality remain reproducible to the extent possible. 

The framework is implemented in Python and available on Github \cite{jaegli_llmevaluationtoolllmgeneratecodeevaluationtool}. All artifacts during the execution of the pipeline are persisted in a centralized SQLite database.   

\textit{Exercise definition:} The programming exercises (tasks or prompts describing what code to write) are stored as version text-files in a git repository and then imported into the prompt table in the database. This allows clear exercise inputs over runs that are identical and documented.

\textit{Code generation:} Automated model invocation with strictly defined input-output parameters ensures a traceable audit trail for all generated code artifacts.  For LLMs, we used OpenAI API with fixed request parameters (client, model, temperature, top p, max tokens, frequency penalty and presence penalty) and defined response parameters (timestamp, output text, response id, request id, finish reason, latency ms, output tokens, total tokens).  The parameters were fixed across models and defined in \ref{AppendixMetrics}. Generated code is captured in a defined output format to minimize variability.  


This step is repeated for a fixed number of times, creating a number of code outputs for each exercise and model, before moving to the next step in the pipeline.  

\textit{Code execution:} To ensure consistent initial conditions, both human and LLM-generated code are executed in isolated Podman containers with a fixed set of local dependencies. We further mitigate environmental non-determinism by restricting tasks to self-contained logic that does not rely on external network services.

\label{security_metrics}
\textit{Code evaluation:} As described in Table \ref{tab:qualitycriteria} , we used a collection of metrics to measure correctness, clarity, and security of the code output.  A rule-based assessment was performed with static Ruff linter \cite{gonzalez_ruff_2024}.   

\begin{table*}
    \centering 
    \caption{Quality Metrics and Criteria}
    \label{tab:qualitycriteria}
    \begin{tabular}{p{0.15\linewidth} p{0.25\linewidth} p{0.5\linewidth}} 
        \toprule
        \textbf{Quality Criteria} & \textbf{Metric} & \textbf{Description} \\ 
        \midrule
        Correctness & Pass (1) / fail (0) & Output of the code execution is evaluated automatically to determine if the output is correct or not. \\ 
        \addlinespace 
        Clarity & Cyclomatic complexity & Measures independent paths through the code sample. High complexity indicates hard-to-understand logic.  \\  
        & Function length & Counts the number of lines in a function. Large functions are harder to understand. \\ 
        \addlinespace
        \bottomrule
    \end{tabular}
\end{table*}

\subsubsection{Reproducibility and Auditing}
The pipeline captures metadata to create traceable artifacts linking specific prompt inputs to quality outputs. Metatdata includes: model parameters, timestamps, execution logs, and security results. This structured environment ensures the reproducibility of deterministic properties, such as initial conditions and cause-effect relationships.

However, while the framework facilitates characterization through repeated trials, absolute predictability remains hindered by the intrinsic non-determinism of LLM internals \cite{ouyang2025empirical} and external service variability. To preserve the integrity of these trials, execution and evaluation results remain inaccessible to the models, preventing the LLMs from learning across multiple iterations and ensuring that each generation is independent.

\subsubsection{Containers for Security and Reproducibility}
We use Podman (an open-source, daemonless container engine) for reproducibility \cite{walsh_podman_2023}, but they serve a secondary purpose in protecting the researchers' environment.  Evaluating code security requires executing potentially malicious or malformed logic.  We avoid these pitfalls by routing all generated code through a rootless Podman containerization layer.




\subsection{Framework Results}
\label{ResultFramework}
The framework implements four primary controls to isolate variables identified in Developer-Centered Security (DCS) literature while adhering to the species-fair doctrine \cite{firestone_performance_2020}.

\subsubsection{Species Fair Integrity}

\textit{Environmental Symmetry:} To ensure fair constraints, we utilize Podman-based containerization. This guarantees that both human and LLM-generated code are evaluated within identical operating systems, library sets, and hardware allocations.

\textit{Instruction Parity:} Functional requirements and security constraints must be mirrored exactly between human instructions and LLM system prompts. This prevents prompt-tuning bias, where one agent receives more architectural guidance than the other.

\textit{Contextual Calibration}: DCS research highlights participant expertise as a critical determinant of security. The framework allows researchers to map model capabilities (e.g., base vs. reasoning models) to commensurate human experience levels (e.g., students vs. seniors).

\textit{Functional Quality Gates}: To avoid the common pitfall of evaluating the security of non-functional code, the pipeline enforces sequential evaluation. Code must pass functional unit tests before undergoing security analysis, ensuring results are not confounded by syntax or logical failures.

\textit{Isolation and Traceability}: Each iteration generates a unique audit trail linking prompts to outcomes. The environment resets for every trial to prevent data leakage or cross-iteration learning, ensuring independent results.

\subsubsection{Research Utility and Framework Benefits}
The framework provides a flexible, extensible foundation for comparative studies across diverse code-generation paradigms.

\textit{Multi-dimensional Assessment}: The pipeline automates the evaluation of execution, code clarity, and security. We leverage the Ruff linter (800+ rules), allowing researchers to toggle security patterns relevant to their specific exercises (e.g., omitting SQL injection checks for mathematical tasks).

\textit{Reproducibility }: By minimizing measurement variance in inherently non-deterministic LLM environments, the framework provides a stable baseline. 

\textit{Operational Efficiency}: Designed for accessibility, the toolchain utilizes ubiquitous technologies (Python, SQLite, Podman). During our feasibility study, OpenAI API costs were minimal (approx. \$10), and the containerized architecture protected the host machine from potentially malicious code execution.

\textit{Privacy}: To protect human participants , the framework is a ``closed loop'' It does not feed code back into LLMs for training or prompts, preventing the proliferation of private information (such as humans including their identifiers in the code).

\section{Feasibility Study}
\label{FSMethods}

To demonstrate the framework's application in assessing both LLM-generated and human-authored code, we executed a feasibility study focused on Python security and algorithmic challenges. 

\subsection{Exercise selection}
Thirteen coding exercises were chosen as prompts. These prompts cover Python security exercises and Advent of Code (AoC) 2024 challenges.\cite{aoc_advent_2025}. These tasks span difficulty levels from beginner to expert and cover diverse problem types (algorithmic puzzles, parsing, data-structure task) as well as security-focused cases. These challenges require both syntactic correctness and algorithmic reasoning, expose security and robustness issues, and allow deterministic validation of correctness via predefined input/expected output pairs.  Where available, property-based unit tests were used to verify and stabilize basic building blocks of the code base \cite{borstler_developers_2023}.  

\subsubsection{Human Code Data sources}
A set of human reference solutions was included in execution and quality assessment steps (see Figure \ref{fig:evaluation_pipeline}). For w3resource exercises, only one online solution was available. For Advent of Code 2024, hundreds of solutions exist on Github.  Approximately five high-ranking Python solutions were selected from the public leader board and extracted from their repositories.  

\begin{wrapfigure}[14]{r}{0.5\textwidth} 
  \vspace{-30pt} 
  \centering
  \includegraphics[width=0.48\textwidth]{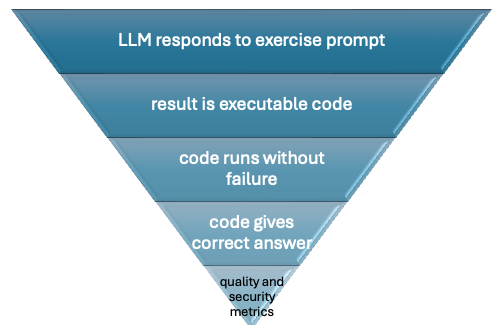}
  \caption{Cascading errors: Results at each step may be disqualified from inclusion in the next analysis.}
  \label{fig:errorsteps}
  \vspace{-10pt} 
\end{wrapfigure}

\subsubsection{LLM Code Generation}
Exercises were copied from the challenge websites and used as prompts. For this process input data are embedded in the prompt text file, as the LLMs cannot load external directories.

Five LLM model variants were compared: gpt-4.1, gpt-4o-mini, gpt-5.1, gpt-5-mini, and gpt-5-nano. According to the OpenAI documentation, these models differ in coding reasoning, determinism, latency, cost per run, error profile or failure modes. \cite{openai_models_2025} 

Each exercise and model combination can executed in the automated pipeline, as described in \ref{fig:evaluation_pipeline}, multiple times. Prompt ingestion, parametrization (temperature, top p, max tokens, frequency penalty and presence penalty), and input token number were stable across runs.  All requests to generate code completed without error.

\subsubsection{Handling Errors}

In this analysis, code is analyzed in multiple steps, as shown in Figure \ref{fig:errorsteps}.  For example, the LLMs did not always output executable code; sometimes they just returned the answer or human text intermingled with code snippets. We did not do further execution or testing on non-code results. In these cases, the sample was removed from further testing. Some of the executable code failed to run, or ran and produced incorrect answers. If it failed, the analysis stopped for that sample. The assessment for cyclomatic complexity, function length and security metrics was done for correct code results only.





\subsection{Feasibility Study Results}
\label{FSResults}
We utilize a pilot sample to examine the nuances of species-fair task constraints and automated security evaluation. Rather than providing definitive security conclusions, these preliminary results stress-test the evaluation pipeline, ensuring the framework can reliably support the complex, species-fair research it was built to facilitate.

\subsubsection{Data Analysis}
Each exercise was generated 60 times per model to minimize within-task variance and approximate independent draws \cite{alvarado_gonzalez_repetitions_2025}.  60 repetitions were intended strike a balance between capturing variability in model output and practical resource and time constraints.  However, as described previously, we were not able to run all samples. In our case, the total test should have 3,900 data points (5 models * 13 exercises  * 60 times), but we were only able to execute 3,832 samples (98\%) of LLM-generated code, and 2,615 produced correct output (67\%).  Furthermore, some models never produced correct code for some exercises.  This leads to a skewed dataset which violates many common distribution assumptions.  Assumptions for linear regression, ANOVA, and poison regression did not hold due to the dispersion and distribution of these errors.  

The selection of top-rated human code samples for this feasibility study introduced an inherent ceiling effect, as all human-authored entries executed without functional errors. These samples are not intended to represent a statistically random distribution of developer performance. Rather, the impact of this selection process reinforces our primary thesis: that future comparative research requires a standardized framework built from the ground up to account for and mitigate such experimental biases.

\subsubsection{Correctness}
For each exercise, the ``correct" answer was a pre-defined string.  We verified whether the executed code generated this correct answer.   Correctness varied greatly across the exercises.  Some exercises were solved consistently across all models, others showed more model-dependent performance.  Correctness was near‑100\% for gpt‑4.1, strong but task‑sensitive for gpt‑4o‑mini, and more variable for the gpt‑5 family (with gpt‑5.1 out-performing gpt‑5‑mini/nano).    

To account for the bounded nature and unequal variance of the correctness scores (0 to 60), we employed a Binomial Generalized Linear Model (GLM) using a logit link function to model the correctness. The relative influence of prompt id and model was then quantified by comparing the increase in model deviance upon the removal of each variable, identifying the factor that contributed most significantly to the model's overall fit. We find the exercise itself was more predictive of correct output than the different models. 

\begin{table}[ht]
\centering
\begin{tabular}{lcccc}
\toprule
\textbf{Predictor} & \textbf{$\Delta$ Deviance} & \textbf{$df$} & \textbf{$p$-value} & \textbf{Influence Rank} \\
\midrule
Prompt ID & 1693.59 & 12 & < 0.001 & 1st \\
Model     & 397.81 & 4 & < 0.001 & 2nd \\
\bottomrule
\end{tabular}
\caption{Analysis of Deviance for Correctness: The higher variance for the prompt id indicates that the exercise had more influence on correctness than the model.}
\label{table:correctness_analysis}
\end{table}

\subsubsection{Failure Mode}
Analysis of failure modes revealed two dominant error types: input/output and structural errors.  (1) “No output” error type meaning that the executed code did not print a correct standard output.  Analysis of the code showed that this error is caused by its improper input handling. The generated solutions assumed external files (e.g., password.txt), expected manual input, omitted input assignment, or printed natural-language answers instead of executable code.  (2) Algorithmic and structure errors: The LLM generated functioning code but the output was wrong ( “WrongLogic”), or the script was aborted due to  ``IndexError”, ``RecursionError” or``NameErrors”. 

\longonly{
  Aggregated error counts of incorrectly solved challenges are shown in ~\ref{fig:security_failure_codes}.
\begin{figure}[h]
    \centering
        \includegraphics[width=1\linewidth]{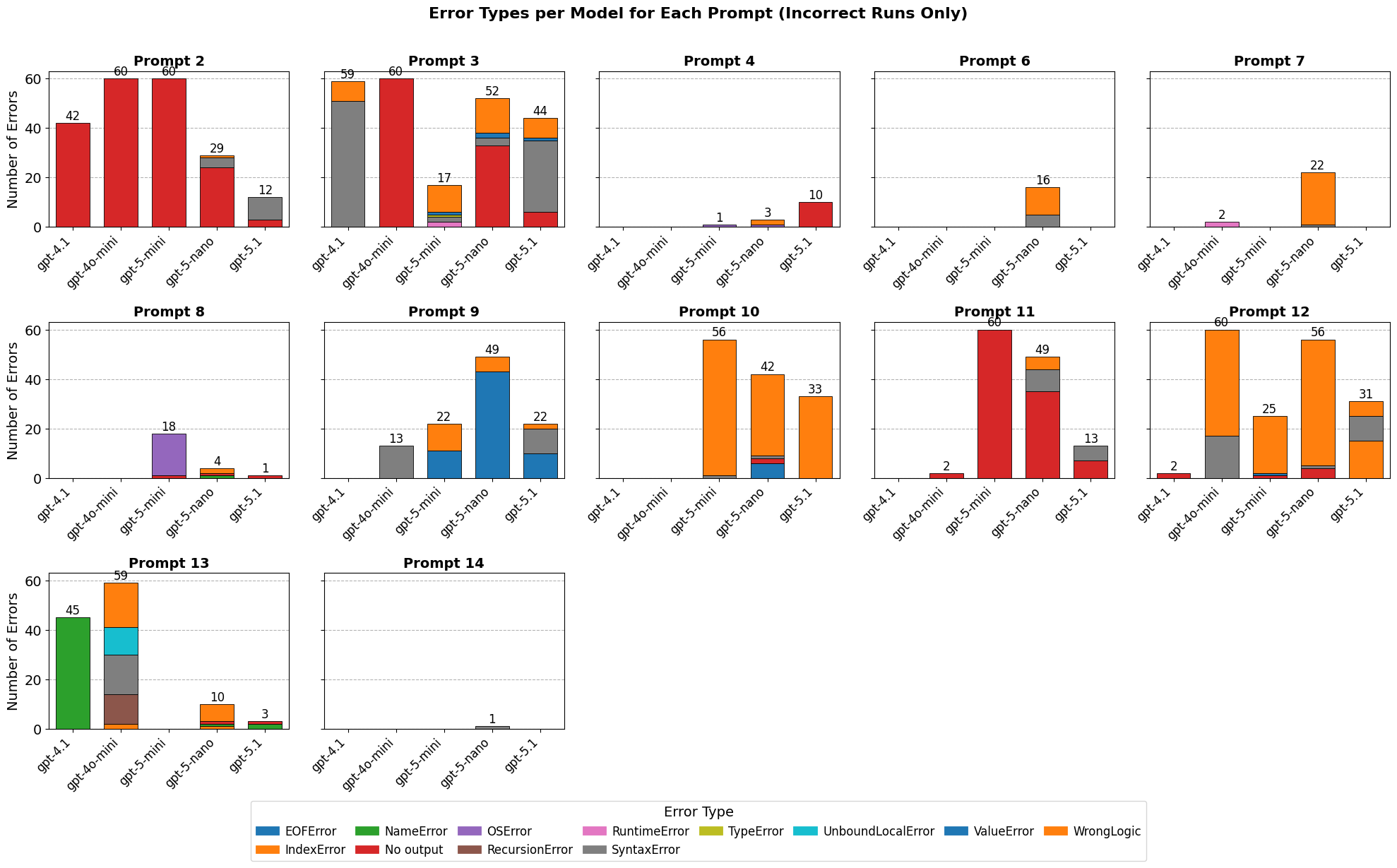} 
        \caption{Failure Modes: The types of errors across models and runs. WrongLogic or No Output were the most common errors. }
        \label{fig:security_failure_codes}
\end{figure}
}

\subsubsection{Clarity} 
Code clarity is characterized by cyclomatic complexity and function length.  Comparative analysis of correctly solved runs highlighted distinct stylistic differences between models and human authors.  See Table \ref{table:complexity-scores}.  Human-authored references generally exhibited lower McCabe cyclomatic complexity than the models, despite the functions being longer. The GPT-4 series produced less complex solutions than the GPT-5 series on harder tasks.  Similar to McCabe cyclomatic complexity, we assessed the quality metric ``function length”. Typical mean values varied from 4 up to 20 lines, with CLT standard errors ranging from ±0.3 to ±2.0 depending on sample size (n). 



\begin{table}[ht]
\begin{tabular}{lllllll}
\toprule
model & gpt-4.1  & gpt-4o-mini  & gpt-5-mini  & gpt-5-nano  & gpt-5.1  & human \\
  &  &  &  &  &  &  \\
\midrule
McCabe & 4.6 ± 0.1 & 3.3 ± 0.0 & 5.2 ± 0.1 & 4.2 ± 0.1 & 4.5 ± 0.1 & 2.9 ± 0.2 \\ 
Function Length & 9.2 ± 0.3  & 9.3 ± 0.1  & 7.4 ± 0.7  & 13.5 ± 0.4  & 6.4 ± 0.1  & 10.7 ± 0.2   \\
\bottomrule
\end{tabular}
\caption{Complexity Scores: Mean value of cyclomatic complexity (McCabe) with CLT standard error for correctly solved challenges only across different models and  human generated references. Mean value of function length with CLT standard error for correctly solved challenges across different models and human generated references.}
\label{table:complexity-scores}
\end{table}

\subsubsection{Security}
The total number of security issues reported was small, limiting the statistical conclusion.  However, the data suggests that security error types were clustered by exercise rather than the model family, and similar errors occurred within an exercise across models.  While the counts were small for the human-authored code, it appears that errors that the models commonly make in an exercise were also seen in the human-generated code.  

\begin{figure}[h]
    \centering
        \includegraphics[width=1\linewidth]{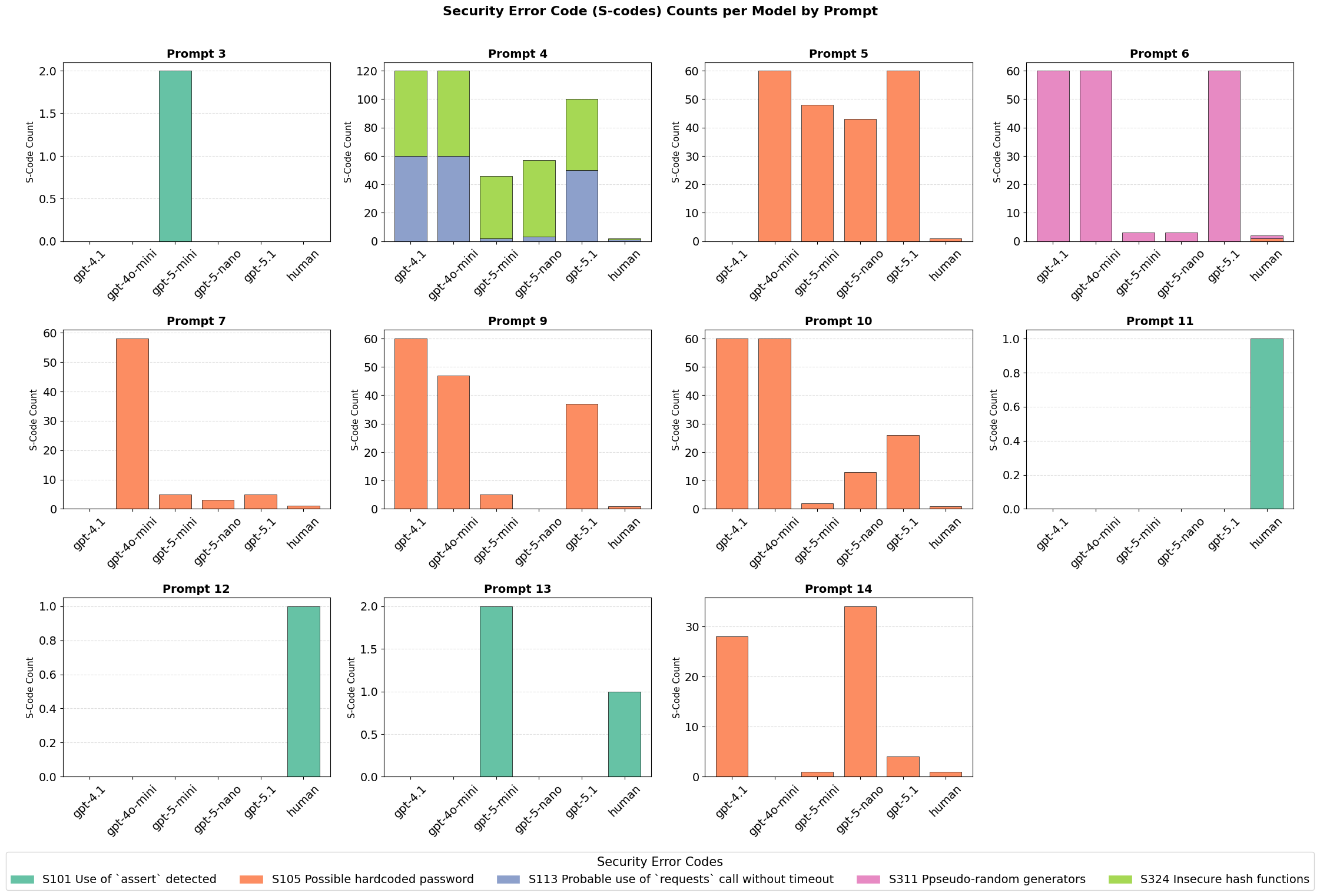} 
        \caption{Security Errors: The types of security errors flagged by Ruff per model and prompt. }
        \label{fig:security_error_codes}
\end{figure}

Figure ~\ref{fig:security_error_codes} summarizes security rule violations across models and human-generated code. No issues were flagged for prompt 2 or 8. 


\section{Discussion}
\label{Discussion}

There is still work to do to understand where humans versus LLMs should be writing code for security. This work does not claim to offer a final answer that one or the other is ``better." Instead, we offer a framework so that more work can be done.  The framework is designed to eliminate confounding factors, ensuring that the results reflect true security differences rather than artifacts of a mismatched experimental design.  

This framework prioritizes a species-fair way to compare code written by humans versus LLM by giving LLMs and humans the same prompts and evaluating them on the same metrics.  Such a prompting style may seem suboptimal for either the humans or the LLMs, but allows for comparable results. In this manner, we can compare apples to apples to better understand strengths and weaknesses of secure code output for similar tasks. 

The feasibility study offers lessons learned for future experiments. The feasibility study can be used to help researchers estimate statistical power or anticipate experimental concerns before comput-time is expended or human subjects are recruited. 

\subsection{Limitations}
The following limitations in the framework were observed:
\begin{itemize}
    \item Task coverage and metric sensitivity: some quality metrics, notably security, did not reach statistical significance for the selected problems. The framework identified that current sample sizes in the literature are likely underpowered. 
    \item While the current reference implementation is Python-specific, the containerized architecture and evaluation logic are language-agnostic.
    \item This work was carried out in late 2025.  The rapid increase in model capability may influence future work.
\end{itemize}

\subsection{Future Work}
Future research should transition toward task-specific, focused experiments designed to identify the exact points in a workflow where LLMs provide the most value and where human intervention remains essential.

\textit{Exercise Selection:} Our results suggest that exercise selection influences outcomes more than model choice, necessitating a focused understanding of how specific tasks elicit vulnerabilities. 

\textit{Prompt Engineering:} Prompt engineering is crucial in providing a fair comparison.  For example, given the exercises instructions and context, humans may understand to write code, but LLMs may need this explicitly stated.  However, researchers must maintain a balance between prompt optimization for LLMs and human readability. 

\textit{Model Selection:} The quality of models overall set a baseline performance. However, increased model reasoning capacity did not inherently translate to higher-quality code.  There are opportunities for future work here. 

\textit{Metric Selection:}
The framework supports a modular selection of metrics tailored to specific research questions.  We report results separately to ensure they remain adaptable to diverse research goals. Developing a single score would require a subjective weighting system that may not align with every study.


\subsection{Ethical Considerations}
Common ethical concerns of this study were discussed by the researchers and considered to be acceptable.  No vulnerabilities in production code or in real systems were discovered in this work. There are no direct military applications. No human subjects were used in this work. The data collection adhered to the terms of service of the respective platforms and follows established norms for the use of publicly available datasets in software engineering research \cite{kohno_ethical_2023}.

\section*{Declaration on Generative AI}
During the preparation of this work, the author(s) used Gemini in order to: Paraphrase and reword, improve writing style, draft code, and simulate peer review.

\bibliography{sections/references}

@inproceedings{astekin_exploratory_2024,
	address = {New York, NY, USA},
	series = {{InteNSE} '24},
	title = {An {Exploratory} {Study} on {How} {Non}-{Determinism} in {Large} {Language} {Models} {Affects} {Log} {Parsing}},
	isbn = {979-8-4007-0564-9},
	doi = {10.1145/3643661.3643952},
	urldate = {2026-01-22},
	booktitle = {Proceedings of the {ACM}/{IEEE} 2nd {International} {Workshop} on {Interpretability}, {Robustness}, and {Benchmarking} in {Neural} {Software} {Engineering}},
	publisher = {Association for Computing Machinery},
	author = {Astekin, Merve and Hort, Max and Moonen, Leon},
	month = aug,
	year = {2024},
	pages = {13--18},
}

@misc{cui_language_2025,
	title = {Do {Language} {Models} {Have} {Bayesian} {Brains}? {Distinguishing} {Stochastic} and {Deterministic} {Decision} {Patterns} within {Large} {Language} {Models}},
	shorttitle = {Do {Language} {Models} {Have} {Bayesian} {Brains}?},
	doi = {10.48550/arXiv.2506.10268},
	urldate = {2026-01-22},
	publisher = {arXiv},
	author = {Cui, Andrea Yaoyun and Yu, Pengfei},
	month = jun,
	year = {2025},
	note = {arXiv:2506.10268 [cs]},
}

@inproceedings{sawadogo_revisiting_2025,
	title = {Revisiting the {Non}-{Determinism} of {Code} {Generation} by the {GPT}-3.5 {Large} {Language} {Model}},
	issn = {2640-7574},
	doi = {10.1109/SANER64311.2025.00012},
	urldate = {2026-01-22},
	booktitle = {2025 {IEEE} {International} {Conference} on {Software} {Analysis}, {Evolution} and {Reengineering} ({SANER})},
	author = {Sawadogo, Salimata and Sabane, Aminata and Kafando, Rodrique and Kabore, Abdoul Kader and Bissyande, Tegawendé F.},
	month = mar,
	year = {2025},
	note = {ISSN: 2640-7574},
	pages = {36--44},
}

@inproceedings{tahaei_survey_2019,
	title = {A {Survey} on {Developer}-{Centred} {Security}},
	url = {https://ieeexplore.ieee.org/document/8802434},
	doi = {10.1109/EuroSPW.2019.00021},
	urldate = {2026-01-22},
	booktitle = {2019 {IEEE} {European} {Symposium} on {Security} and {Privacy} {Workshops} ({EuroS}\&{PW})},
	author = {Tahaei, Mohammad and Vaniea, Kami},
	month = jun,
	year = {2019},
	pages = {129--138},
}

@inproceedings{gutfleisch_how_2022,
	title = {How {Does} {Usable} {Security} ({Not}) {End} {Up} in {Software} {Products}? {Results} {From} a {Qualitative} {Interview} {Study}},
	issn = {2375-1207},
	shorttitle = {How {Does} {Usable} {Security} ({Not}) {End} {Up} in {Software} {Products}?},
	url = {https://ieeexplore.ieee.org/document/9833756},
	doi = {10.1109/SP46214.2022.9833756},
	urldate = {2026-01-22},
	booktitle = {2022 {IEEE} {Symposium} on {Security} and {Privacy} ({SP})},
	author = {Gutfleisch, Marco and Klemmer, Jan H. and Busch, Niklas and Acar, Yasemin and Sasse, M. Angela and Fahl, Sascha},
	month = may,
	year = {2022},
	note = {ISSN: 2375-1207},
	pages = {893--910},
}

@inproceedings{balebako_privacy_2014,
	address = {San Diego, CA},
	title = {The {Privacy} and {Security} {Behaviors} of {Smartphone} {App} {Developers}},
	isbn = {978-1-891562-37-2},
	doi = {10.14722/usec.2014.23006},
	language = {en},
	urldate = {2026-01-22},
	booktitle = {Proceedings 2014 {Workshop} on {Usable} {Security}},
	publisher = {Internet Society},
	author = {Balebako, Rebecca and Marsh, Abigail and Lin, Jialiu and Hong, Jason and Faith Cranor, Lorrie},
	year = {2014},
}

@inproceedings{stransky_lessons_2017,
	title = {Lessons {Learned} from {Using} an {Online} {Platform} to {Conduct} {Large}-{Scale}, {Online} {Controlled} {Security} {Experiments} with {Software} {Developers}},
	url = {https://www.usenix.org/conference/cset17/workshop-program/presentation/stransky},
	language = {en},
	urldate = {2026-01-22},
	author = {Stransky, Christian and Acar, Yasemin and Nguyen, Duc Cuong and Wermke, Dominik and Kim, Doowon and Redmiles, Elissa M. and Backes, Michael and Garfinkel, Simson and Mazurek, Michelle L. and Fahl, Sascha},
	year = {2017},
}

@inproceedings{acar_you_2016,
	address = {Boston, MA, USA},
	title = {You are {Not} {Your} {Developer}, {Either}: {A} {Research} {Agenda} for {Usable} {Security} and {Privacy} {Research} {Beyond} {End} {Users}},
	isbn = {978-1-5090-5589-0},
	shorttitle = {You are {Not} {Your} {Developer}, {Either}},
	url = {http://ieeexplore.ieee.org/document/7839782/},
	doi = {10.1109/SecDev.2016.013},
	language = {en},
	urldate = {2026-01-22},
	booktitle = {2016 {IEEE} {Cybersecurity} {Development} ({SecDev})},
	publisher = {IEEE},
	author = {Acar, Yasemin and Fahl, Sascha and Mazurek, Michelle L.},
	month = nov,
	year = {2016},
	pages = {3--8},
}

@inproceedings{braz_software_2022,
	address = {New York, NY, USA},
	series = {{ESEC}/{FSE} 2022},
	title = {Software security during modern code review: the developer’s perspective},
	isbn = {978-1-4503-9413-0},
	shorttitle = {Software security during modern code review},
	url = {https://dl.acm.org/doi/10.1145/3540250.3549135},
	doi = {10.1145/3540250.3549135},
	urldate = {2026-01-22},
	booktitle = {Proceedings of the 30th {ACM} {Joint} {European} {Software} {Engineering} {Conference} and {Symposium} on the {Foundations} of {Software} {Engineering}},
	publisher = {Association for Computing Machinery},
	author = {Braz, Larissa and Bacchelli, Alberto},
	month = nov,
	year = {2022},
	pages = {810--821},
}

@inproceedings{meli_how_2019,
	address = {San Diego, CA},
	title = {How {Bad} {Can} {It} {Git}? {Characterizing} {Secret} {Leakage} in {Public} {GitHub} {Repositories}},
	isbn = {978-1-891562-55-6},
	shorttitle = {How {Bad} {Can} {It} {Git}?},
	language = {en},
	urldate = {2026-01-22},
	booktitle = {Proceedings 2019 {Network} and {Distributed} {System} {Security} {Symposium}},
	publisher = {Internet Society},
	author = {Meli, Michael and McNiece, Matthew R. and Reaves, Bradley},
	year = {2019},
}

@inproceedings{fischer_stack_2017,
	title = {Stack {Overflow} {Considered} {Harmful}? {The} {Impact} of {Copy}\&{Paste} on {Android} {Application} {Security}},
	issn = {2375-1207},
	shorttitle = {Stack {Overflow} {Considered} {Harmful}?},
	url = {https://ieeexplore.ieee.org/abstract/document/7958574},
	doi = {10.1109/SP.2017.31},
	urldate = {2026-01-22},
	booktitle = {2017 {IEEE} {Symposium} on {Security} and {Privacy} ({SP})},
	author = {Fischer, Felix and Böttinger, Konstantin and Xiao, Huang and Stransky, Christian and Acar, Yasemin and Backes, Michael and Fahl, Sascha},
	month = may,
	year = {2017},
	note = {ISSN: 2375-1207},
	pages = {121--136},
}

@inproceedings{acar_comparing_2017,
	title = {Comparing the {Usability} of {Cryptographic} {APIs}},
	issn = {2375-1207},
	url = {https://ieeexplore.ieee.org/abstract/document/7958576},
	doi = {10.1109/SP.2017.52},
	urldate = {2026-01-24},
	booktitle = {2017 {IEEE} {Symposium} on {Security} and {Privacy} ({SP})},
	author = {Acar, Yasemin and Backes, Michael and Fahl, Sascha and Garfinkel, Simson and Kim, Doowon and Mazurek, Michelle L. and Stransky, Christian},
	month = may,
	year = {2017},
	note = {ISSN: 2375-1207},
	pages = {154--171},
}

@inproceedings{assal_think_2019,
	address = {New York, NY, USA},
	series = {{CHI} '19},
	title = {'{Think} secure from the beginning': {A} {Survey} with {Software} {Developers}},
	isbn = {978-1-4503-5970-2},
	shorttitle = {'{Think} secure from the beginning'},
	url = {https://dl.acm.org/doi/10.1145/3290605.3300519},
	doi = {10.1145/3290605.3300519},
	urldate = {2026-01-26},
	booktitle = {Proceedings of the 2019 {CHI} {Conference} on {Human} {Factors} in {Computing} {Systems}},
	publisher = {Association for Computing Machinery},
	author = {Assal, Hala and Chiasson, Sonia},
	month = may,
	year = {2019},
	pages = {1--13},
}

@inproceedings{kaur_where_2022,
	title = {Where to {Recruit} for {Security} {Development} {Studies}: {Comparing} {Six} {Software} {Developer} {Samples}},
	isbn = {978-1-939133-31-1},
	shorttitle = {Where to {Recruit} for {Security} {Development} {Studies}},
	url = {https://www.usenix.org/conference/usenixsecurity22/presentation/kaur},
	language = {en},
	urldate = {2026-01-26},
	author = {Kaur, Harjot and Klivan, Sabrina and Votipka, Daniel and Acar, Yasemin and Fahl, Sascha},
	year = {2022},
	pages = {4041--4058},
}

@inproceedings{wang_end-users_2024,
	title = {End-{Users} vs {Software} {Practitioners}: {Recruitment} {Challenges} and {Strategies} in {Software} {Engineering} {Research}},
	issn = {1943-6106},
	shorttitle = {End-{Users} vs {Software} {Practitioners}},
	url = {https://ieeexplore.ieee.org/abstract/document/10714544},
	doi = {10.1109/VL/HCC60511.2024.00063},
	urldate = {2026-01-26},
	booktitle = {2024 {IEEE} {Symposium} on {Visual} {Languages} and {Human}-{Centric} {Computing} ({VL}/{HCC})},
	author = {Wang, Wei and Hidellaarachchi, Dulaji and Grundy, John and Khalajzadeh, Hourieh and Obie, Humphrey O. and Madugalla, Anuradha},
	month = sep,
	year = {2024},
	note = {ISSN: 1943-6106},
	pages = {400--411},
}

@inproceedings{alami_are_2024,
	address = {New York, NY, USA},
	series = {{WSESE} '24},
	title = {Are {You} a {Real} {Software} {Engineer}? {Best} {Practices} in {Online} {Recruitment} for {Software} {Engineering} {Studies}},
	isbn = {979-8-4007-0567-0},
	shorttitle = {Are {You} a {Real} {Software} {Engineer}?},
	url = {https://dl.acm.org/doi/10.1145/3643664.3648207},
	doi = {10.1145/3643664.3648207},
	urldate = {2026-01-26},
	booktitle = {Proceedings of the 1st {IEEE}/{ACM} {International} {Workshop} on {Methodological} {Issues} with {Empirical} {Studies} in {Software} {Engineering}},
	publisher = {Association for Computing Machinery},
	author = {Alami, Adam and Zahedi, Mansooreh and Ernst, Neil},
	month = aug,
	year = {2024},
	pages = {52--57},
}

@inproceedings{tahaei_lessons_2022,
	title = {Lessons {Learned} {From} {Recruiting} {Participants} {With} {Programming} {Skills} for {Empirical} {Privacy} and {Security} {Studies}: {RoPES} - {ICSE} 2022},
	shorttitle = {Lessons {Learned} {From} {Recruiting} {Participants} {With} {Programming} {Skills} for {Empirical} {Privacy} and {Security} {Studies}},
	url = {https://ropes-workshops.github.io/ropes22/},
	urldate = {2026-01-26},
	author = {Tahaei, Mohammad and Vaniea, Kami},
	month = may,
	year = {2022},
}

@inproceedings{serafini_recruitment_2023,
	title = {On the {Recruitment} of {Company} {Developers} for {Security} {Studies}: {Results} from a {Qualitative} {Interview} {Study}},
	isbn = {978-1-939133-36-6},
	shorttitle = {On the {Recruitment} of {Company} {Developers} for {Security} {Studies}},
	url = {https://www.usenix.org/conference/soups2023/presentation/serafini},
	language = {en},
	urldate = {2026-01-26},
	author = {Serafini, Raphael and Gutfleisch, Marco and Horstmann, Stefan Albert and Naiakshina, Alena},
	year = {2023},
	pages = {321--340},
}

@article{charoenwet_toward_2024,
	title = {Toward effective secure code reviews: an empirical study of security-related coding weaknesses},
	volume = {29},
	issn = {1573-7616},
	shorttitle = {Toward effective secure code reviews},
	url = {https://doi.org/10.1007/s10664-024-10496-y},
	doi = {10.1007/s10664-024-10496-y},
	language = {en},
	number = {4},
	urldate = {2026-01-26},
	journal = {Empirical Software Engineering},
	author = {Charoenwet, Wachiraphan and Thongtanunam, Patanamon and Pham, Van-Thuan and Treude, Christoph},
	month = jun,
	year = {2024},
	pages = {88},
}

@inproceedings{sandoval_lost_2023,
	title = {Lost at {C}: {A} {User} {Study} on the {Security} {Implications} of {Large} {Language} {Model} {Code} {Assistants}},
	isbn = {978-1-939133-37-3},
	shorttitle = {Lost at {C}},
	url = {https://www.usenix.org/conference/usenixsecurity23/presentation/sandoval},
	language = {en},
	urldate = {2026-01-26},
	author = {Sandoval, Gustavo and Pearce, Hammond and Nys, Teo and Karri, Ramesh and Garg, Siddharth and Dolan-Gavitt, Brendan},
	year = {2023},
	pages = {2205--2222},
}

@article{manes_art_2021,
	title = {The {Art}, {Science}, and {Engineering} of {Fuzzing}: {A} {Survey}},
	volume = {47},
	issn = {1939-3520},
	shorttitle = {The {Art}, {Science}, and {Engineering} of {Fuzzing}},
	url = {https://ieeexplore.ieee.org/abstract/document/8863940},
	doi = {10.1109/TSE.2019.2946563},
	number = {11},
	urldate = {2026-01-26},
	journal = {IEEE Transactions on Software Engineering},
	author = {Manès, Valentin J.M. and Han, HyungSeok and Han, Choongwoo and Cha, Sang Kil and Egele, Manuel and Schwartz, Edward J. and Woo, Maverick},
	month = nov,
	year = {2021},
	pages = {2312--2331},
}

@inproceedings{shin_empirical_2008,
	address = {New York, NY, USA},
	series = {{ESEM} '08},
	title = {An empirical model to predict security vulnerabilities using code complexity metrics},
	isbn = {978-1-59593-971-5},
	url = {https://dl.acm.org/doi/10.1145/1414004.1414065},
	doi = {10.1145/1414004.1414065},
	urldate = {2026-01-26},
	booktitle = {Proceedings of the {Second} {ACM}-{IEEE} international symposium on {Empirical} software engineering and measurement},
	publisher = {Association for Computing Machinery},
	author = {Shin, Yonghee and Williams, Laurie},
	month = oct,
	year = {2008},
	pages = {315--317},
}

@inproceedings{edmundson_empirical_2013,
	address = {Berlin, Heidelberg},
	title = {An {Empirical} {Study} on the {Effectiveness} of {Security} {Code} {Review}},
	isbn = {978-3-642-36563-8},
	doi = {10.1007/978-3-642-36563-8_14},
	language = {en},
	booktitle = {Engineering {Secure} {Software} and {Systems}},
	publisher = {Springer},
	author = {Edmundson, Anne and Holtkamp, Brian and Rivera, Emanuel and Finifter, Matthew and Mettler, Adrian and Wagner, David},
	editor = {Jürjens, Jan and Livshits, Benjamin and Scandariato, Riccardo},
	year = {2013},
	pages = {197--212},
}

@article{belozerov_secure_2025,
	title = {Secure {Coding} with {AI}, {From} {Creation} to {Inspection}},
	journal = {arXiv preprint arXiv:2504.20814},
	author = {Belozerov, Vladislav and Barclay, Peter J and Sami, Ashkan},
	year = {2025},
}

@inproceedings{perry_users_2023,
	title = {Do users write more insecure code with ai assistants?},
	booktitle = {Proceedings of the 2023 {ACM} {SIGSAC} conference on computer and communications security},
	author = {Perry, Neil and Srivastava, Megha and Kumar, Deepak and Boneh, Dan},
	year = {2023},
	pages = {2785--2799},
}

@article{becker_measuring_2025,
	title = {Measuring the impact of early-2025 {AI} on experienced open-source developer productivity},
	journal = {arXiv preprint arXiv:2507.09089},
	author = {Becker, Joel and Rush, Nate and Barnes, Elizabeth and Rein, David},
	year = {2025},
}

@misc{openai_createtranslation_nodate,
	title = {{createTranslation}},
	url = {https://platform.openai.com/docs/api-reference/audio/createTranslation#audio_createtranslation-temperature},
	urldate = {2025-09-21},
	author = {{OpenAI}},
}

@article{akamine_effects_2024,
	title = {Effects of temperature settings on information quality of {ChatGPT}-3.5 responses: {A} prospective, single-blind, observational cohort study},
	journal = {medRxiv},
	publisher = {Cold Spring Harbor Laboratory Press},
	author = {Akamine, Akihiko and Hayashi, Daisuke and Tomizawa, Atsushi and Nagasaki, Yuya and Akamine, Chikae and Fukawa, Takahiro and Hirosawa, Iori and Saigo, Orie and Hayashi, Misa and Nanaoya, Mitsuru},
	year = {2024},
	pages = {2024.06. 11.24308759},
}

@misc{openai_create_nodate,
	title = {Create chat completion},
	url = {https://platform.openai.com/docs/api-reference/chat/create#chat_create-frequency_penalty},
	language = {English},
	urldate = {2025-09-21},
	author = {{OpenAI}},
}

@book{walsh_podman_2023,
	title = {Podman in {Action}: {Secure}, rootless containers for {Kubernetes}, microservices, and more},
	isbn = {1-63343-968-2},
	publisher = {Simon and Schuster},
	author = {Walsh, Daniel},
	year = {2023},
}

@misc{verdier_fastest_2023,
	title = {The {Fastest} {Way} to {Boost} your {Code} {Quality}: {Use} {Ruff} {Linter}},
	url = {https://data-ai.theodo.com/en/technical-blog/boost-code-quality-ruff-linter},
	language = {English},
	author = {Verdier, Edmont},
	month = jun,
	year = {2023},
}

@misc{gonzalez_ruff_2024,
	title = {Ruff: a game changer or {Python} linters.},
	url = {https://xantygc.medium.com/ruff-a-game-changer-for-python-linters-12b1ec8c5f12},
	author = {Gonzalez, Santiage},
	month = nov,
	year = {2024},
}

@article{brtnik_analysis_nodate,
	title = {Analysis and {Extension} of the {Ruff} {Linter}},
	author = {BRTNÍK, TOMÁŠ},
}

@techreport{kristinsson_implementing_2024,
	title = {Implementing {Python} code quality checks in the {CMSSW} {Continuous} {Integration} infrastructure},
	author = {Kristinsson, Benedikt Tor},
	year = {2024},
}

@article{borstler_developers_2023,
	title = {Developers talking about code quality},
	volume = {28},
	number = {6},
	journal = {Empirical Software Engineering},
	publisher = {Springer},
	author = {Börstler, Jürgen and Bennin, Kwabena E. and Hooshangi, Sara and Jeuring, Johan and Keuning, Hieke and Kleiner, Carsten and MacKellar, Bonnie and Duran, Rodrigo and Störrle, Harald and Toll, Daniel},
	year = {2023},
	pages = {128},
}

@inproceedings{gosain_static_2015,
	title = {Static analysis: {A} survey of techniques and tools},
	booktitle = {Intelligent {Computing} and {Applications}: {Proceedings} of the {International} {Conference} on {ICA}, 22-24 {December} 2014},
	publisher = {Springer},
	author = {Gosain, Anjana and Sharma, Ganga},
	year = {2015},
	pages = {581--591},
}

@inproceedings{choi_static_2021,
	address = {Cham},
	title = {Static {Analysis} for {Software} {Reliability} and {Security}},
	isbn = {978-3-030-71017-0},
	booktitle = {Advances in {Security}, {Networks}, and {Internet} of {Things}},
	publisher = {Springer International Publishing},
	author = {Choi, Hongjun and Kang, Dayoung and Choi, Jin-Young},
	editor = {Daimi, Kevin and Arabnia, Hamid R. and Deligiannidis, Leonidas and Hwang, Min-Shiang and Tinetti, Fernando G.},
	year = {2021},
	pages = {463--470},
}

@misc{openai_codex_2025,
	title = {Codex},
	url = {https://openai.com/de-DE/codex/},
	language = {EN},
	author = {{OpenAI}},
	month = nov,
	year = {2025},
}

@misc{noauthor_w3resource_2024,
	author = {{w3resource}},
	title = {w3resource},
	url = {https://www.w3resource.com/python-exercises/cybersecurity/},
	month = nov,
	year = {2024},
}

@misc{aoc_advent_2025,
	title = {Advent of {Code} 2024},
	url = {https://adventofcode.com/2024/},
	author = {{AoC}},
	month = nov,
	year = {2025},
}

@misc{aoc_advent_2025-1,
	title = {Advent of {Code} 2025},
	url = {https://adventofcode.com/2025/},
	language = {EN},
	author = {{AoC}},
	month = feb,
	year = {2025},
}

@misc{openai_models_2025,
	title = {Models},
	url = {https://platform.openai.com/docs/models},
	urldate = {2025-12-02},
	author = {{OpenAI}},
	month = feb,
	year = {2025},
}

@article{alvarado_gonzalez_repetitions_2025,
	title = {Do {Repetitions} {Matter}? {Strengthening} {Reliability} in {LLM} {Evaluations}},
	journal = {arXiv e-prints},
	author = {Alvarado Gonzalez, Miguel Angel and Hernandez, Michelle Bruno and Peñaloza Perez, Miguel Angel and Lopez Orozco, Bruno and Tadeo Cruz Soto, Jesus and Malagon, Sandra},
	year = {2025},
	pages = {arXiv: 2509.24086},
}

@article{pistoia_survey_2007,
	title = {A survey of static analysis methods for identifying security vulnerabilities in software systems},
	volume = {46},
	number = {2},
	journal = {IBM systems journal},
	publisher = {IBM},
	author = {Pistoia, Marco and Chandra, Satish and Fink, Stephen J. and Yahav, Eran},
	year = {2007},
	pages = {265--288},
}

@inproceedings{marvin_prompt_2024,
	address = {Singapore},
	title = {Prompt {Engineering} in {Large} {Language} {Models}},
	isbn = {978-981-99-7962-2},
	doi = {10.1007/978-981-99-7962-2_30},
	language = {en},
	booktitle = {Data {Intelligence} and {Cognitive} {Informatics}},
	publisher = {Springer Nature},
	author = {Marvin, Ggaliwango and Hellen, Nakayiza and Jjingo, Daudi and Nakatumba-Nabende, Joyce},
	editor = {Jacob, I. Jeena and Piramuthu, Selwyn and Falkowski-Gilski, Przemyslaw},
	year = {2024},
	pages = {387--402},
}

@inproceedings{bruni_benchmarking_2025,
	title = {Benchmarking {Prompt} {Engineering} {Techniques} for {Secure} {Code} {Generation} with {GPT} {Models}},
	url = {https://ieeexplore.ieee.org/abstract/document/11052790},
	doi = {10.1109/Forge66646.2025.00018},
	urldate = {2026-02-03},
	booktitle = {2025 {IEEE}/{ACM} {Second} {International} {Conference} on {AI} {Foundation} {Models} and {Software} {Engineering} ({Forge})},
	author = {Bruni, Marc and Gabrielli, Fabio and Ghafari, Mohammad and Kropp, Martin},
	month = apr,
	year = {2025},
	pages = {93--103},
}

@inproceedings{song2025good,
  title={The good, the bad, and the greedy: Evaluation of llms should not ignore non-determinism},
  author={Song, Yifan and Wang, Guoyin and Li, Sujian and Lin, Bill Yuchen},
  booktitle={Proceedings of the 2025 Conference of the Nations of the Americas Chapter of the Association for Computational Linguistics: Human Language Technologies (Volume 1: Long Papers)},
  pages={4195--4206},
  year={2025}
}

@article{beckers2025large,
  title={Large Language Models as Nondeterministic Causal Models},
  author={Beckers, Sander},
  journal={arXiv preprint arXiv:2509.22297},
  year={2025}
}

@article{ouyang2025empirical,
  title={An empirical study of the non-determinism of chatgpt in code generation},
  author={Ouyang, Shuyin and Zhang, Jie M and Harman, Mark and Wang, Meng},
  journal={ACM Transactions on Software Engineering and Methodology},
  volume={34},
  number={2},
  pages={1--28},
  year={2025},
  publisher={ACM New York, NY}
}

@inproceedings{shin2008empirical,
  title={An empirical model to predict security vulnerabilities using code complexity metrics},
  author={Shin, Yonghee and Williams, Laurie},
  booktitle={Proceedings of the Second ACM-IEEE international symposium on Empirical software engineering and measurement},
  pages={315--317},
  year={2008}
}

@article{moshtari2013using,
  title={Using complexity metrics to improve software security},
  author={Moshtari, Sara and Sami, Ashkan and Azimi, Mahdi},
  journal={Computer Fraud \& Security},
  volume={2013},
  number={5},
  pages={8--17},
  year={2013},
  publisher={Elsevier}
}

@article{tehrani2024assessing,
  title={Assessing Vulnerability in Smart Contracts: The Role of Code Complexity Metrics in Security Analysis},
  author={Tehrani, Masoud Jamshidiyan and Hashemi, Sattar},
  journal={arXiv preprint arXiv:2411.17343},
  year={2024}
}

@misc{dai_rethinking_2025,
    title = {Rethinking the {Evaluation} of {Secure} {Code} {Generation}},
    doi = {10.48550/arXiv.2503.15554},
    urldate = {2026-02-16},
    publisher = {arXiv},
    author = {Dai, Shih-Chieh and Xu, Jun and Tao, Guanhong},
    month = nov,
    year = {2025},
    note = {arXiv:2503.15554 [cs]},
}

@inproceedings{he_large_2023,
    address = {New York, NY, USA},
    series = {{CCS} '23},
    title = {Large {Language} {Models} for {Code}: {Security} {Hardening} and {Adversarial} {Testing}},
    isbn = {979-8-4007-0050-7},
    shorttitle = {Large {Language} {Models} for {Code}},
    url = {https://dl.acm.org/doi/10.1145/3576915.3623175},
    doi = {10.1145/3576915.3623175},
    urldate = {2026-02-18},
    booktitle = {Proceedings of the 2023 {ACM} {SIGSAC} {Conference} on {Computer} and {Communications} {Security}},
    publisher = {Association for Computing Machinery},
    author = {He, Jingxuan and Vechev, Martin},
    month = nov,
    year = {2023},
    pages = {1865--1879},
}

@inproceedings{he_instruction_2024,
    address = {Vienna, Austria},
    series = {{ICML}'24},
    title = {Instruction tuning for secure code generation},
    volume = {235},
    urldate = {2026-02-18},
    booktitle = {Proceedings of the 41st {International} {Conference} on {Machine} {Learning}},
    publisher = {JMLR.org},
    author = {He, Jingxuan and Vero, Mark and Krasnopolska, Gabriela and Vechev, Martin},
    month = jul,
    year = {2024},
    pages = {18043--18062},
}

@misc{nijkamp_codegen_2023,
    title = {{CodeGen}: {An} {Open} {Large} {Language} {Model} for {Code} with {Multi}-{Turn} {Program} {Synthesis}},
    shorttitle = {{CodeGen}},
    url = {http://arxiv.org/abs/2203.13474},
    doi = {10.48550/arXiv.2203.13474},
    urldate = {2026-02-18},
    publisher = {arXiv},
    author = {Nijkamp, Erik and Pang, Bo and Hayashi, Hiroaki and Tu, Lifu and Wang, Huan and Zhou, Yingbo and Savarese, Silvio and Xiong, Caiming},
    month = feb,
    year = {2023},
    note = {arXiv:2203.13474 [cs]},
}

@article{firestone_performance_2020,
	title = {Performance vs. competence in human–machine comparisons},
	volume = {117},
	issn = {0027-8424, 1091-6490},
	doi = {10.1073/pnas.1905334117},
	language = {en},
	number = {43},
	urldate = {2026-02-05},
	journal = {Proceedings of the National Academy of Sciences},
	author = {Firestone, Chaz},
	month = oct,
	year = {2020},
	pages = {26562--26571},
}

@inproceedings{kohno_ethical_2023,
	title = {Ethical {Frameworks} and {Computer} {Security} {Trolley} {Problems}: {Foundations} for {Conversations}},
	isbn = {978-1-939133-37-3},
	shorttitle = {Ethical {Frameworks} and {Computer} {Security} {Trolley} {Problems}},
	url = {https://www.usenix.org/conference/usenixsecurity23/presentation/kohno},
	language = {en},
	urldate = {2026-01-29},
	author = {Kohno, Tadayoshi and Acar, Yasemin and Loh, Wulf},
	year = {2023},
	pages = {5145--5162},
    booktitle = {Usenix Security}
}

@inproceedings{ramirez_state_2024,
    title = {State of the {Art} of the {Security} of {Code} {Generated} by {LLMs}: {A} {Systematic} {Literature} {Review}},
    shorttitle = {State of the {Art} of the {Security} of {Code} {Generated} by {LLMs}},
    doi = {10.1109/CONISOFT63288.2024.00050},
    urldate = {2025-10-14},
    booktitle = {2024 12th {International} {Conference} in {Software} {Engineering} {Research} and {Innovation} ({CONISOFT} 2024)},
    author = {Ramírez, Leonardo Criollo and Limón, Xavier and  Sánchez-García, Angel J. and Pérez-Arriaga, Juan Carlos},
    month = oct,
    year = {2024},
    pages = {331--339},
}

@misc{licorish_comparing_2025,
    title = {Comparing {Human} and {LLM} {Generated} {Code}: {The} {Jury} is {Still} {Out}!},
    shorttitle = {Comparing {Human} and {LLM} {Generated} {Code}},
    doi = {10.48550/arXiv.2501.16857},
    urldate = {2026-03-08},
    publisher = {arXiv},
    author = {Licorish, Sherlock A. and Bajpai, Ansh and Arora, Chetan and Wang, Fanyu and Tantithamthavorn, Kla},
    month = jan,
    year = {2025},
    note = {arXiv:2501.16857 [cs]},
}

@misc{molison_is_2025,
    title = {Is {LLM}-{Generated} {Code} {More} {Maintainable} \& {Reliable} than {Human}-{Written} {Code}?},
    doi = {10.48550/arXiv.2508.00700},
    urldate = {2026-03-08},
    publisher = {arXiv},
    author = {Molison, Alfred Santa and Moraes, Marcia and Melo, Glaucia and Santos, Fabio and Assuncao, Wesley K. G.},
    month = aug,
    year = {2025},
    note = {arXiv:2508.00700 [cs]},
}

@inproceedings{cotroneo_human-written_2025,
    title = {Human-{Written} vs. {AI}-{Generated} {Code}: {A} {Large}-{Scale} {Study} of {Defects}, {Vulnerabilities}, and {Complexity}},
    issn = {2332-6549},
    shorttitle = {Human-{Written} vs. {AI}-{Generated} {Code}},
    doi = {10.1109/ISSRE66568.2025.00035},
    urldate = {2026-03-08},
    booktitle = {2025 {IEEE} 36th {International} {Symposium} on {Software} {Reliability} {Engineering} ({ISSRE})},
    author = {Cotroneo, Domenico and Improta, Cristina and Liguori, Pietro},
    month = oct,
    year = {2025},
    note = {ISSN: 2332-6549},
    pages = {252--263},
}

@misc{jaegli_llmevaluationtoolllmgeneratecodeevaluationtool,
	title = {{LLMEvaluationTool}/{LLMgenerateCodeEvaluationTool} at main · jaegli/{LLMEvaluationTool}},
	url = {https://github.com/jaegli/LLMEvaluationTool/},
	language = {en},
    year={2026},
	urldate = {2026-05-28},
	journal = {GitHub},
	author = {{Egli, Jasmine}},
}

@inproceedings{ortloff_sok_2023,
    title = {{SoK}: {I} {Have} the ({Developer}) {Power}! {Sample} {Size} {Estimation} for {Fisher}'s {Exact}, {Chi}-{Squared}, {McNemar}'s, {Wilcoxon} {Rank}-{Sum}, {Wilcoxon} {Signed}-{Rank} and t-tests in {Developer}-{Centered} {Usable} {Security}},
    isbn = {978-1-939133-36-6},
    shorttitle = {{SoK}},
    url = {https://www.usenix.org/conference/soups2023/presentation/ortloff},
    language = {en},
    urldate = {2025-09-08},
    author = {Ortloff, Anna-Marie and Tiefenau, Christian and Smith, Matthew},
    year = {2023},
    pages = {341--359},
}

\longonly{
\appendix
\section{Appendix: Framework and Metrics}
\label{AppendixMetrics}


Execution and static analysis are coordinated by the analyse\_outputs routine, which launches each script in its own container, executes it and performs static checks. With a minimal monitoring system, the analyse\_outputs orchestration routine verifies required configurations (e.g., ruff\_cfg, output\_directory), executes all Python scripts in the output directory, logs execution details and records per-run metadata (started\_at, exit\_code, standard output (stdout), standard error (stderr), and duration\_ms)in the podman\_runs table.  

\begin{table}[ht]
\caption{Error Type Distribution Across All Models and Prompts in Feasibility Study}
\label{tab:total_errors}
\begin{tabular}{l|r|r}
\toprule
Error Type & Count & Percentage (\%) \\
\midrule
No output & 426 & 35.4 \\
WrongLogic & 394 & 32.7 \\
SyntaxError & 194 & 16.1 \\
EOFError & 70 & 5.8 \\
NameError & 49 & 4.1 \\
OSError & 19 & 1.6 \\
IndexError & 18 & 1.5 \\
RecursionError & 12 & 1.0 \\
UnboundLocalError & 11 & 0.9 \\
ValueError & 6 & 0.5 \\
RuntimeError & 4 & 0.3 \\
TypeError & 1 & 0.1 \\
\textbf{Total} & 1204 & 100.0 \\
\bottomrule
\end{tabular}
\label{table:complexity}
\end{table}

\subsection{Ruff Linter Security Codes}
For automated code-quality assessment, a linter in combination with custom functions were used. Linters detect syntax errors, security anti-patterns, or style violations\cite{verdier_fastest_2023}. After evaluating Pylint, Pyflakes, Pycodestyle, Flake8, Bandit, and Ruff, Ruff, a Rust-based linter, was selected for its execution speed and extensive rule set (800 rules, including Bandit security checks). \cite{verdier_fastest_2023,brtnik_analysis_nodate,gonzalez_ruff_2024,kristinsson_implementing_2024} Ruff was configured via pyproject.toml, mounted into rootless Podman containers, and executed through the lint\_script\_in\_container function. Ruff findings and the custom metrics were recorded in the ruff\_issues table; The analyse\_outputs routine orchestrates logging, database connections, Podman volume mounts, script execution, and linting (including custom metric calculation). 

The Ruff Security Codes discussed in this paper are as follows:
\begin{itemize}
    \item S101 (assert used): Use of the assert statement. While common in testing, it is flagged in production code because assert statements can be optimized away when Python is run with the optimize flag, potentially bypassing critical security checks.
    \item S105 (hardcoded password string): Possible hardcoded password or credential. This is triggered when the linter finds a string assignment to a variable name that looks like a secret (e.g., password = "12345").
    \item S113 (request without timeout): Probable use of a requests call without a timeout parameter. This prevents your application from hanging indefinitely if a server fails to respond.
    \item S311 (suspicious non cryptographic random usage): Use of the standard random library for security or cryptographic purposes. The random module is pseudo-random and predictable; for security (like tokens or passwords), secrets or os.urandom should be used instead.
    \item S324 (hashlib insecure hash function): Use of insecure hash functions like MD5 or SHA1 for sensitive data. These functions are vulnerable to collision attacks and should be replaced with stronger algorithms.
\end{itemize}

\section{Appendix: LLM Information}
\label{AppendixLLM}

Five LLM model variants were compared: gpt-4.1, gpt-4o-mini, gpt-5.1, gpt-5-mini, and gpt-5-nano. According to the OpenAI documentation\cite{openai_codex_2025}. These models differ in coding reasoning, determinism, latency, cost per run, error profile or failure modes. The following claims are made about the models: gpt-5.1 and gpt-5-mini offer the highest coding and reasoning quality  (but at higher cost and latency), while gpt-5-nano and gpt-4o-mini prioritize very low cost and fast throughput with reduced reasoning depth. gpt-4.1 provides a stable, more deterministic baseline useful for reproducible comparisons between stochastic, higher-capability and low-cost models.\cite{openai_models_2025}.

\begin{table}[h]
\centering
\begin{tabular}{ l | l | l | l | l | l}
model & gpt-4.1  & gpt-4o-mini  & gpt-5-mini  & gpt-5-nano  & gpt-5.1  \\
date &  2025-04-14 &  2024-07-18 &  2025-08-07 &  2025-08-07 &  2025-11-13 \\
\end{tabular}
\caption{Model names and dates}
\label{tab:modelinformation}
\end{table}

Execution and static analysis are coordinated by the analyse\_outputs routine, which launches each script in its own container, executes it and performs static checks. With a minimal monitoring system, the analyse\_outputs orchestration routine verifies required configurations (e.g., ruff\_cfg, output\_directory), executes all Python scripts in the output directory, logs execution details and records per-run metadata \textemdash started\_at, exit\_code, standard output (stdout), standard error (stderr), and duration\_ms \textemdash in the podman\_runs table.

\begin{table}[h]
\centering
\begin{tabularx}{\textwidth}{ l | l | X } 
\hline
\textbf{Request parameter} & \textbf{Set value} & \textbf{Description}  \\
client & OpenAI & Official OpenAI HTTP client used to send requests and receive responses and sets uniform client configuration.  \\
model & \textit{e.g}. gpt-4.1 & LLM variants used for a given generation.  \\
temperature & 0.0 & To minimize randomness. A value of 0 supports determinism. \cite{akamine_effects_2024}  \\
top\_p & 1.0 & Nucleus sampling cutoff; 1.0 disables top-p filtering so all tokens are considered.\cite{openai_createtranslation_nodate} \\
max\_tokens & 15000 & Max. number of response tokens.  \\
frequency\_penalty & 0.0 & Penalizes repeated tokens; set to 0 to avoid altering model repetition behavior.\cite{openai_create_nodate}  \\
presence\_penalty & 0.0 & Penalizes introducing new topics; set to 0 to avoid discouraging content variation.\cite{openai_create_nodate}   \\
\end{tabularx}
\caption{To achieve maximal determinism,\textsuperscript{[24]}\textsuperscript{[25]}\textsuperscript{[26]} the input variables were controlled and the generations were issued via the OpenAI API with fixed request parameters.}
\label{tab:parameterinformation}
\end{table}

\begin{table}[h]
\centering
\begin{tabularx}{\textwidth}{ l | l | X } 
\hline
\textbf{Response parameter} & \textbf{Value} & \textbf{Description} \\
stop\_reason & stop,length, timeout, error. & Reason the API reports why generation ended: ‘stop’ indicates normal completion.   \\
request\_id & uid
  & Unique identifier for the API request (useful for tracing and support).  \\
response\_id & uid & Unique identifier for the specific response object returned by the API (links request to output). \\
created\_at & timestamp & Timestamp when the response was created by the API server   \\
Latency\_ms & in milliseconds & Custom: Round-trip time in milliseconds between sending the request and receiving the response; used to monitor performance and variability.  \\
\end{tabularx}
\caption{In order to achieve maximal determinism, the output variables from the generation via OpenAI API were defined as response parameters.}
\label{tab:parameterinformation2}
\end{table}


\section{Appendix: Coding Exercises}
\label{AppendixExercises}

The coding challenges are taken from the following resources: w3resource\cite{noauthor_w3resource_2024}, Advent of Code 2024\cite{aoc_advent_2025}, and Advent of Code 2025\cite{aoc_advent_2025-1}.  They are described in Table \ref{tab:cybersecurity_exercises}.  

\begin{table}[ht]
\centering
\small
\renewcommand{\arraystretch}{1.5}
\begin{tabularx}{\textwidth}{| l | >{\raggedright\arraybackslash}p{2.5cm} | >{\raggedright\arraybackslash}X | >{\raggedright\arraybackslash}X |}
\hline
\textbf{prompt\_id} & \textbf{Exercise name} & \textbf{Complexity level} & \textbf{Link to original exercise \& code} \\
\hline
2 & AoC 2024 Day 17 & Advanced & \href{https://github.com/FelixCenusa/Advent-Of-Code-2024/blob/main/Day-17-Challenge/day17p1.py}{GitHub: FelixCenusa} \\
\hline
3 & AoC 2024 Day 21 & Advanced & \href{https://github.com/FelixCenusa/Advent-Of-Code-2024/blob/main/Day-21-Challenge/day21p1.py}{GitHub: FelixCenusa} \\
\hline
4 & Leaked Password Checker & Intermediate: Involves external API integration, including handling HTTP requests and responses. Requires understanding data formats (e.g., JSON) and dealing with potential edge cases, making this exercise significantly more complex. & \href{https://www.w3resource.com/python-exercises/cybersecurity/python-cybersecurity-exercise-6.php}{w3resource Ex 6} \\
\hline
5 & SHA-256 Hasher & Intermediate: Requires understanding hashing concepts, managing input/output in Python, and utilizing the hashlib library. Implementation is straightforward but requires attention to detail. & \href{https://www.w3resource.com/python-exercises/cybersecurity/python-cybersecurity-exercise-1.php}{w3resource Ex 1} \\
\hline
6 & Random Password Generators & Basic: Involves random number generation and string manipulation. & \href{https://www.w3resource.com/python-exercises/cybersecurity/python-cybersecurity-exercise-2.php}{w3resource Ex 2} \\
\hline
7 & Common Substitution Generator & Intermediate: Requires an understanding of character mapping and the implications of substitutions in password strength. & \href{https://www.w3resource.com/python-exercises/cybersecurity/python-cybersecurity-exercise-4.php}{w3resource Ex 4} \\
\hline
8 & Password File Generator & Involves file I/O and validation logic. & \href{https://www.w3resource.com/python-exercises/cybersecurity/python-cybersecurity-exercise-5.php}{w3resource Ex 5} \\
\hline
9 & Password strength meter & Basic & \href{https://www.w3resource.com/python-exercises/cybersecurity/python-cybersecurity-exercise-7.php}{w3resource Ex 7} \\
\hline
10 & Brute Force Attack & Intermediate & \href{https://www.w3resource.com/python-exercises/cybersecurity/python-cybersecurity-exercise-10.php}{w3resource Ex 10} \\
\hline
11 & AoC 2024 Day 1 & Advanced: requires understanding both sorting list and iterating through lists, along with basic arithmetic. & \href{https://github.com/FelixCenusa/Advent-Of-Code-2024/blob/main/Day-11-Challenge/day11p1.py}{GitHub: Solution11a}, \href{https://github.com/jamiemilsom/adventOfCode-2024/blob/main/day1/day1.py}{GitHub: Solution11b}, \href{https://github.com/Hamatti/adventofcode-2024/blob/main/src/day_1.py}{GitHub: Solution11c}, \href{https://github.com/purpleysound/advent-of-code/blob/main/2024/python_oneliners/day_1.py}{GitHub: Solution11d} \\
\hline
12 & AoC 2024 Day 8 & Advanced: Understanding of programming skills (mathematical calculations, nested loops, conditions checks, edge cases as well as geometric concepts like antinodes). & \href{https://github.com/FelixCenusa/Advent-Of-Code-2024/blob/main/Day-08-Challenge/day8p1.py}{GitHub: Solution12a}, \href{https://github.com/obhalerao/AoC2024/blob/main/day08/Day08A.py}{GitHub: Solution12b}, \href{https://github.com/fugleder/adventofcode/blob/master/2024/day08/solutions.py}{GitHub: Solution12c} \\
\hline
13 & AoC 2025 Day 10 & Advanced: implementing graph traversal algorithm or similar methods. & \href{https://github.com/FelixCenusa/Advent-Of-Code-2024/blob/main/Day-10-Challenge/day10p1.py}{GitHub: Solution13a}, \href{https://github.com/Bikatr7/AdventOfCode/blob/main/solutions/2024/10/10.py}{GitHub: Solution13d}, \href{https://github.com/iKevinY/advent/blob/main/2024/day10.py}{GitHub: Solution13e} \\
\hline
14 & Password Criteria Checker & Basic: Involves conditional statements and string validation techniques. Requires systematic logic to check multiple criteria, making it slightly more challenging than basic syntax. & \href{https://www.w3resource.com/python-exercises/cybersecurity/python-cybersecurity-exercise-3.php}{w3resource Ex 3} \\
\hline
16 & AoC 2025 Day 1 & Advanced & No reference \\
\hline
17 & AoC 2025 Day 2 & Advanced & No reference \\
\hline
\end{tabularx}
\caption{Summary of Cybersecurity and Advent of Code (AoC) programming exercises, categorized by complexity and source.}
\label{tab:cybersecurity_exercises}
\end{table}
}

\end{document}